\documentclass[journal]{IEEEtran}

\usepackage{amsmath,amssymb}
\usepackage{graphicx}
\usepackage[noadjust]{cite}
\usepackage{etoolbox}      
\usepackage{hyperref}

\graphicspath{{./}{fig/}}

\patchcmd{\bordermatrix}{8.75}{4.75}{}{}
\patchcmd{\bordermatrix}{\left(}{\left[}{}{}
\patchcmd{\bordermatrix}{\right)}{\right]}{}{}

\newcommand{\tblhdr}[2]{\hline \multicolumn{#1}{l}{%
 \vbox{ \vspace*{4pt} \hbox{\text{#2}} \vspace*{0pt}} %
 }\\ \hline\rule{0pt}{2.6ex}}

\newcommand{\rpkg}[1]{{\em #1}}

\begin{document}

\title{Runs and Bootstrap Tests For Signal Feature Significance}
\author{Greg Kreider
  \thanks{G. Kreider is with Primordial Machine Vision Systems and can be
          contacted at gkreider@primachvis.com} }

\maketitle

\begin{abstract}
~~Runs tests have long been used as a non-parametric check if data contains
a non-random signal.  We derive a recursive expression for the
distribution of the longest run using Markov chain theory.  Next we
develop a permutation test on the runs comprising a feature to get the
probability of its height. This leads finally to a bootstrap test on
the height using the raw, continuous data.  Such a test can evaluate not
only the large heights of peaks but also the small heights of flats.  We
can apply these tests to features in the spacing of data to detect and
locate multi-modality.
\end{abstract}


While investigating the use of spacing to detect multi-modality, we
needed tests of the probability of features found in it.  Spacing is
the difference between consecutive order statistics, or after sorting.
In univariate draws it has a long flat minimum at the mode
and increases rapidly in the tails.  For multi-variate setups these
tails merge into local peaks at the anti-modes.  Although an integral
for the expected spacing is known, it can be solved for only a few
single distributions and not for any combination.  The tests must
use only the actual data to evaluate the height and duration of a
peak, or the length of a flat that stays within some ripple bounds.

Most signal processing analysis depends on modeling the signal and noise,
from which one can build density functions for a feature and thresholds
for its detection.  Single outlier points, for example, follow directly
from the data's distribution, and extended features can be handled
similarly \cite{helstrom95}.  Distribution functions are available for
the height and length of Brownian processes
\cite{hall97, chung76, iafrate19} but the signal may not have the
necessary Gaussian noise.  The spacing does not.

We will start by considering quantized or binned versions of the data,
which reduces the measured values to a small number of discrete levels.
The expected number of runs within a symbol follows from a combinatorial
analysis.  We will create a Markov chain model of the runs and use it to
test the longest. We will then map runs to the height of a feature and
create a permutation test to estimate its significance.  Finally we will
return to the original data and perform a similar bootstrap test.

The key results include the probability of the longest run
\eqref{eq:pmaxlen} using \eqref{eq:plento} and \eqref{eq:recur}, the
permutation test of runs for features in discrete data in Listing~1, and
a bootstrap test of features in general data in Listing~2.  The last two
are simple extensions or applications of known techniques.

\section{Runs Tests}

Non-parametric tests are available after quantizing the signal, for
example by considering only the sign of the difference between
successive points.  Wallis and Moore \cite{wallis41tr} calculated the
distribution of the lengths of increasing or decreasing series of
continuous values, without ties, using a modified chi-squared test.
This approach has been used for process control \cite{wolfowitz43}.
It is rather coarse, however, stopping at runs of length 3 because
the distribution drops off rapidly.  A combinatorial counting of runs
can handle more than two symbols \cite{mood40, shaughnessy81}.

To avoid the problem of low counts we can use summary tests.  By a
combinatorial analysis of the symbols Kaplansky and Riordan show
the number of runs is normally distributed
\cite[(12) and (13)]{kaplansky45}.  This extends the two symbol
analysis of \cite[(12) and (13)]{wald40}.  \cite{philippou86}
instead gives the probability of the longest run in one of two
symbols.

\subsection{Longest Run Length}

A combinatorial analysis does not include correlations between symbols,
such as those introduced by filtering.  A Markov chain model can, easily
handling an arbitrary symbol set.  We will derive the distribution of
the longest run in general.  Let the data have $ m $ symbols and an
$ m \times m $ transition matrix $ {\bf T} $, using
\cite[(2.8)]{anderson57} to estimate the rates by counting the
transitions out of each state and normalizing each row to sum to 1.  If
the chain has an order $ r $ greater than one \cite{dorea14}, then use
all $ m^r $ symbol combinations.  We build a new transition matrix
$ {\bf R} $ whose states will be the steps of a run.  Split $ {\bf T} $
in two, with $ {\bf A} $ the diagonal elements that advance the run
length within a symbol and $ {\bf B} $ the off-diagonal elements that
move the length back to 1, so that $ {\bf T} = {\bf A} + {\bf B} $.
Each row of $ {\bf R} $ will have $ {\bf B} $ in the first column,
$ {\bf A} $ in the column corresponding to the next longest run, and
0 elsewhere.
\begin{equation} \label{eq:trunlen}
{\bf R} =
 \bordermatrix{ %
 \overset{\text{next}}{\text{\scriptsize length}} & \text{\scriptsize 1} &
   \text{\scriptsize 2} & \text{\scriptsize 3} & \text{\scriptsize L} \cr
 & {\bf B} & {\bf A} & 0 & 0 \cr
 & {\bf B} & 0 & {\bf A} & 0 \cr
 & {\bf B} & 0 & 0 & {\bf A} \cr
 & 0       & 0 & 0 & {\bf I} }
\end{equation}
The identity matrix $ {\bf I} $ is an absorbing state that represents
all runs longer than length $ L $.  $ {\bf R} $ has size
$ (L+1) \times (L+1) $ of $ m \times m $ sub-matrices.  The probability
of a run is
\begin{align} \label{eq:pmaxlen}
P\{\text{run length} \leq L\} & = 1 - P\{\text{absorbed at } L+1\} \nonumber \\
P\{\text{longest run} = L\} & = 
  P\{\text{len} \leq L\} - P\{\text{len} \leq L-1\}
\end{align}
This can be seen as an extension of \cite[Theorem 3.1]{fu94}, with the
matrices $ {\bf A} $ and $ {\bf B} $ replacing single state rates in
order to track runs in all possible states at the same time.

The probability of absorption in a feature of size $ N $ is found in the
top right element $ {\bf r}_{1,L+1,N} = {\bf R}^{N}[1,L+1] $, letting
the Markov chain evolve normally over the $ N $ steps.  We look at only
the first row, ignoring any run in progress at the start of the feature.
The last column is
\begin{equation} \label{eq:lastcol}
{\bf R}^N = \left.
\begin{array}{cc}
  & {\bf B} ~ {\bf r}_{1,L+1,N-1} + {\bf A} ~ {\bf r}_{2,L+1,N-1} \\
  & {\bf B} ~ {\bf r}_{1,L+1,N-1} + {\bf A} ~ {\bf r}_{3,L+1,N-1} \\
\ldots  & \vdots \\
  & {\bf B} ~ {\bf r}_{1,L+1,N-1} + {\bf A} ~ {\bf r}_{l,L+1,N-1} \\
  & {\bf B} ~ {\bf r}_{1,L+1,N-1} + {\bf A} \\
  & {\bf I}
\end{array}
\right]
\end{equation}
Because all column indices are $ L + 1 $, it references only itself.  In
general for row $ i $,
\begin{equation} \label{eq:recur}
{\bf r}_{i,L+1,N} = \left\{
\begin{aligned}
{\bf B} ~ {\bf r}_{1,L+1,N-1} + {\bf A} ~ {\bf r}_{i+1,L+1,N-1} & & i \leq L \\
{\bf I} & & i = L+1
\end{aligned}
\right.
\end{equation}
seeded by
\begin{equation*} \label{eq:recurst}
{\bf r}_{i,L+1,0} = \left\{
\begin{aligned}
0 & & i \leq L \\
{\bf I} & & i = L + 1
\end{aligned}
\right.
\end{equation*}
For $ N \leq L $ the last column fills with powers of $ {\bf A} $,
\begin{equation} \label{eq:colfill}
{\bf r}_{i,L+1,N} = {\bf A} ^ {i - 1}
\end{equation}
letting $ {\bf A} ^ 0 = {\bf I } $.  After this the first column contributes
$ {\bf B} $ times the previous upper right value to the next step, and the
$ {\bf A} $ above the diagonal multiply the previous value in the next row.

The final probability of absorption is found by summing over all states
in $ {\bf r}_{1,L+1,N} $ after weighting by an initial distribution
$ {\bf w} $, an $ m \times 1 $ vector.  The stationary state, or left
eigenvectors, of $ {\bf T} $ gives this distribution on average, or
$ {\bf w} $ can be set to 1 in the position corresponding to a known
starting state and 0 elsewhere. Then
\begin{equation} \label{eq:plento}
P\{\text{run length} \leq L\} = {\bf w} ~ {\bf r}_{1,L+1,N}
\end{equation}
The top right element also obeys the recursion
\begin{equation} \label{eq:recur1}
{\bf r}_{1,L+1,N} =
  {\bf A}^{L} + \sum_{j=1}^{L} {\bf A}^{j-1} ~ {\bf B} ~ {\bf r}_{1,L+1,N-j}
\end{equation}
subject to $ {\bf r}_{1,L+1,n} = 0 $ if $ n < L $ and $ {\bf A} $ if
$ n = L $.  Raising $ {\bf A} $ to a power is easily done because it is
diagonal, but this form requires more storage than \eqref{eq:recur} by
referring to $ N $ instead of $ L $ values, and more matrix
multiplications.

\begin{figure}
\centering
\includegraphics{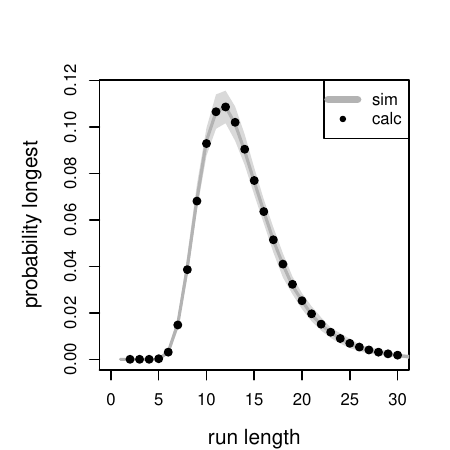}
\caption{\label{fig:maxlen} Verification of the longest run distribution.}
\end{figure}

\figurename~\ref{fig:maxlen} plots the distribution of the longest run
length for data that contains 3 states, in a chain formed by taking the
signed difference of 400 points drawn uniformly between 0 and 1 and
rounded to multiples of 0.02 to create ties.  From its transition
matrix we simulate 2500 trials of sequences of 100 points and tally the
counts of the longest runs.  The grey band in the figure is a 95\%
confidence interval over the trials.  The calculated results come from
\eqref{eq:recur} and fall within the band.

A similar technique can give us the distribution of run lengths.  The
run transition matrix is changed by deleting the absorbing state in
\eqref{eq:trunlen} and continuing both the rows and columns indefinitely.
After $ N $ steps the first $ N $ columns of $ {\bf R}^N $ are constant,
independent of row $ i $ and in column $ l $
\begin{equation}
{\bf r}_{i,l,N} = {\bf R}^{N}[i,l] =
  ({\bf B} + {\bf A})^{N-l} ~ {\bf B} ~ {\bf A}^{l-1}
\end{equation}
The distribution for length $ l $ follows these values,
weighted by $ {\bf w} $.  Since
$ ({\bf B} + {\bf A})^{N-l} = {\bf T}^{N-l} $ quickly becomes constant,
the distribution becomes geometric in $ {\bf A} $.  The expected counts
decrease rapidly enough to make testing against observed counts
difficult, however, as happened with the Wallis and Moore test.

\section{Feature Tests}

\subsection{Run Length Permutation Test}

We can replace the signal by its signed runs, the run length between
each change in slope.  The reconstructed signal, formed by their sum,
simplifies the original while generally preserving its features.  Raw
data with differences of roughly the same magnitude will suffer the
least distortion. One can identify features in either, like peaks
between two local minima, and analyze them in terms of the runs they
contain.

\begin{figure}
\centering
\includegraphics{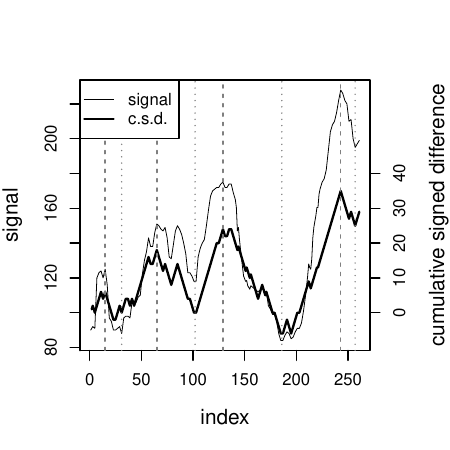}
\caption{\label{fig:cumsign} Raw signal and simplified cumulative signed
  difference (c.s.d.).}
\end{figure}

Consider an artificial raw signal and its reconstruction, the cumulative
signed difference c.s.d., in \figurename~\ref{fig:cumsign} and the
features listed in Table~1.  Dotted vertical lines at local minima bound
peaks marked with dashed lines. One can see that the reconstructed version
damps large steps, decreasing peak heights but generally preserving
their form.  The first two peaks become less distinct.  The positions
of the extrema are stable, except for the first minimum at index 32,
which shifts to the left.  We can identify two heights for a feature:
the start-to-finish difference between the minima, and the largest
difference over the range.  The start-to-finish height comes from the
sum of the signed run lengths and is the same no matter their order.

The peak height, on the other hand, does depend on the pattern of the
runs.  It will be largest if long rising and falling sections lie on
different sides of the maximum, interrupted by only short moves in the
other direction. If the alternating moves have similar lengths,
however, they will tend to cancel, leaving a smaller overall height.
The third peak separates rising from falling runs and has a large
height.  The fourth also separates them but has fewer falling runs and
does not rise much above its start-to-finish height.  The second has a
dip at index 81, which limits its height, and the first mixes its runs.


\begin{center}
{\small
\begin{tabular}{lcccc}
 \\
\tblhdr{5}{Table 1: Run Features Example}%
indices                 & 1--30  & 31--101     & 102--185    & 186--256    \\
peak at                 & 15     & 65          & 129         & 243         \\
start-finish height     & 0      & 0           & -6          & 31          \\
peak height             & 8      & 18          & 30          & 41          \\
$ h_{p} $ range         & 0--8   & 0--20       & 6--30       & 9--39       \\
$ p_{ht} = 1 - q_{ht} $ & 0.070  & {\em 0.027} & {\em 0.004} & {\em 0.002} \\
 & & & & \\
longest run             & 8      & 14          & 20          & 22          \\
$ p_{run} $             & 0.119  & 0.088       & {\em 0.036} & {\em 0.020} \\
\hline
 & & & & \\
\end{tabular}
}
\end{center}

Given a set of runs that comprise a feature, its quantile is the
fraction of reconstructed heights over all permutations that are
smaller (Listing~1).  To ignored inverted excursions we define the
simulated height as the maximum value above the start or finish.
We sample the permutations, with $ N_{perm} = 5000 $ giving stable
results. The only condition placed on them (in step 4a) is that no
two runs in the same symbol may be adjacent, or they would merge
into a longer run which does not exist in the data.  The
implementation of Listing~1 does the permutation in two stages,
first ordering the symbols while maintaining the separation
requirement, then shuffling the lengths within each symbol into place.
If there are two classes the first step is easy, an alternating
sequence with the more frequent symbol first.  With three or more
classes symbols are picked according to their frequency while avoiding
the previous choice, until one has the majority.  It must then
alternate with a shuffle of the remaining classes.  As shown in
Table~1, this test finds the first peak to be unremarkable with a
probability of 0.070, the second better defined at the 0.05 level,
and the other two significant.  They also pass the longest run test.
Notice that the range of simulated heights for the fourth peak does
not reach the actual height.

The distribution of the permuted heights $ h_{p} $ skews to larger
values with a long tail, behaving like a Brownian excursion.
Significant deviations are possible if the peak has side lobes or
level sections.  However, the integer heights and small range of
$ h_{p} $, seen in Table~1, mean that the quantiles will be coarse.
\cite{ma11} uses piecewise linear segments to smooth a discrete
distribution, which amounts to counting half the matching heights
(step 5).  This improves the accuracy of $ q_{ht} $ but does not
remove the steps between different heights.  The resolution of
$ q_{ht} $ near the significance level may be poor.

\begin{center}
{\small
\begin{tabular}{ll}
 \\
\tblhdr{2}{Listing 1:  Height Run Permutation Test}%
1:  & convert signal $ x[1 \ldots n] $ to signed differences \\
    & \quad $ \Delta = \text{sign}( x_{i} - x_{i-1} ) $ \\
2:  & identify feature height $ h_{f} = \text{max}(x) - \text{min}(x) $ \\
3:  & build set of run lengths in $ \Delta $ per sign (-1, 0, +1) \\
4:  & repeat $N_{perm}$ times: \\
4a: & \quad permute run lengths, no same sign adjacent \\
4b: & \quad construct permuted signal $ s $ \\
    & \quad \quad = cumulative sum of lengths $ \times $ sign \\
4c: & \quad height $ h_{p} = \text{max}(s) - \text{min}(s[1], s[n]) $ \\
5:  & $ q_{ht} = (\#(h_{p} < h_{f}) + \#(h_{p} = h_{f}) / 2) / N_{perm} $ \\
\hline
 \\
\end{tabular}
}
\end{center}

\subsection{Bootstrap Test}

\begin{figure}
\centering
\includegraphics{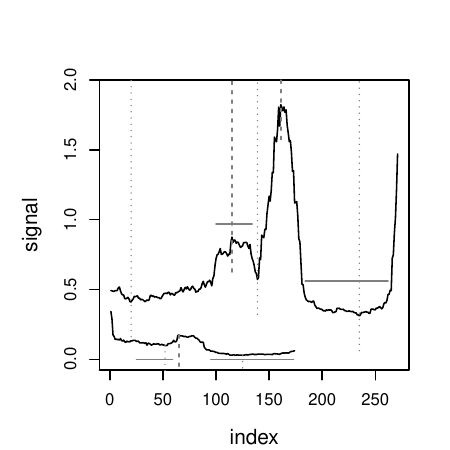}
\caption{\label{fig:diffsig} Two signals, the top with large differences and
the bottom not.}
\end{figure}

This technique also works using the raw differences between data
points as the building blocks.  As with the runs test, this version
depends on separating large differences in opposite directions,
although it does not require alternating steps.
\figurename~\ref{fig:diffsig} shows two artificial examples built
from the spacing of tri- and bi-modal setups.  The
top signal has a mix of steps of all sizes, while the bottom has
only a few moderate changes other than the initial drop.  The
prominent peak at top separates rising and falling steps while the
secondary peak to its left has many small mixed steps in its flat
top, as does the bottom feature.  Because the flat minima
artificially lengthen the top peaks, we take them to extend to
90\% of their height, similar to Full Width at Half Maximum (FWHM).

A permutation of the raw differences within a feature will generate
smaller quantiles because the steps do not represent all the data.
A bootstrap, sampling with replacement from the step differences in
the entire signal (Listing 2), produces more moderate simulated heights
and more conservative probabilities than the permutation test.  On
draws from a uniform distribution, for example, the bootstrap test
finds peaks at the 0.05 level in 4.5\% of trials, while a permutation
test has a high false positive rate of 18.8\%.  To be clear, in our
application the signal $ x $ in Listing~2 is the spacing and the
signed differences $ \Delta $ its derivative.

\begin{center}
{\small
\begin{tabular}{ll}
 \\
\tblhdr{2}{Listing 2: Height Step Bootstrap Test}%
1:  & build set of signal differences within data \\
    & \quad $ \Delta = x_{i} - x_{i-1} $ \\
2:  & identify feature with $ n $ points, height $ h_{f} $ \\
3:  & repeat $ N_{boot} $ times: \\
3a: & \quad sample $ n $ points from $ \Delta $, with replacement \\
3b: & \quad bootstrap signal $ s $ = cumulative sum of sample \\
3c: & \quad height $ h_{b} = \text{max}(s) - \text{min}(s[1], s[n]) $ \\
4:  & $ q_{ht} = ( \#(h_{b} < h_{f}) + \#(h_{b} = h_{f}) / 2) / N_{boot} $ \\
\hline
 \\
\end{tabular}
}
\end{center}

Table~2 summarizes the permutation and bootstrap test results on the
features in \figurename~\ref{fig:diffsig}.  The main peaks in each
signal are significant in the bootstrap test at the 0.05 level, but
the left top peak is not.  Only the main peak in the lower trace
passes the runs permutation test because the signed runs magnify the
slow decline to right minimum.

\begin{center}
{\small
\begin{tabular}{lcccc}
 \\
\tblhdr{5}{Table 2: Excursion Features Example}%
trace              & top, left   & top, right  & bottom &    \\
indices            & 20--139     & 139--235    & 52--142     \\
peak at            & 115         & 161         & 65          \\
s-f height         & 0.161       & -0.260      & -0.069      \\
peak height        & 0.464       & 1.512       & 0.145       \\
$ h_{b} $ range    & 0.06--1.98  & 0.05--1.77  & 0.01--0.18  \\
$ p_{ht,boot} $    & 0.700       & {\em 0.016} & {\em 0.024} \\
 & & & & \\
runs height        & 21          & 24          & 32          \\
$ h_{p} $ range    & 7--25       & 11--26      & 18--31      \\
$ p_{ht,perm} $    & 0.074       & 0.078       & {\em 0.001} \\
\hline
 & & & & \\
\end{tabular}
}
\end{center}

The bootstrap test applies to any features characterized by their height
and size.  An example would be flats, defined as a section whose height
remains within a ripple specification, here 10\% of the signal.
Figure~\ref{fig:diffsig} marks four flats with horizontal lines.  The
simulated height is now the range $ h_{b} = \max(s) - \min(s) $ to match
the ripple and the quantile counts the trials greater than $ h_{b} $.
The upper example has a level section around the right minima that is
significant (Table~3) and one at the top of the left peak that is
marginal.  The bottom left flat is not significant but the right is.
These judgements reflect the flat's length and the signal's roughness.

\begin{center}
{\small
\begin{tabular}{lcccc}
 \\
\tblhdr{5}{Table 3: Flat Features Example}%
trace             & top         & top         & bottom      & bottom       \\
indices           & 100--134    & 184--262    & 25--59      & 95--173      \\
length            & 35          & 79          & 35          & 79           \\
height            & 0.155       & 0.152       & 0.032       & 0.030        \\
$ h_{b} $ range   & 0.08--1.03  & 0.22--1.58  & 0.01--0.12  & 0.03--0.19   \\
$ p_{flat} $      & 0.059       & {\em 0.0}   & 0.259       & {\em 0.010}  \\
\hline
 & & & & \\
\end{tabular}
}
\end{center}

\section{Software and Application}
These tests are used in \rpkg{Dimodal}, an R package for analyzing
modality using spacing \cite{kreider25c}.  It provides the reference
implementation of the tests described here.  It performs the runs tests
on the interval spacing, which uses larger spans of the order statistics
and acts as a rectangular filter on the raw spacing; the poor sidelobe
repression of this filter means the interval spacing will be rough and
will support runs.  The bootstrap test is performed on the spacing
after low-pass filtering, which is essential to reduce the variance.
The sample pool must exclude the largest spacings found in the tails,
above three standard deviations, as these differences are not
representative of those making up the features.

\begin{figure}
\centering
\includegraphics{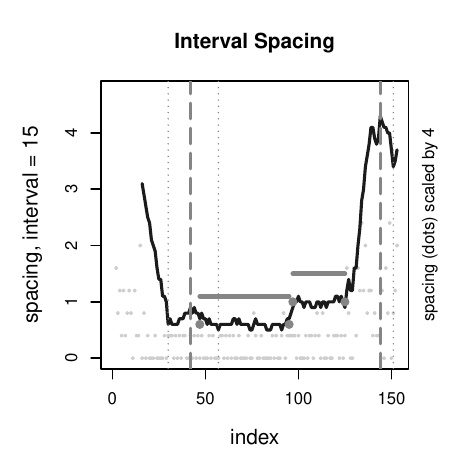}
\caption{\label{fig:maas} Spacing analysis of organic matter density near the
 Maas river. }
\end{figure}

As an example, we look at soil data collected near the Maas River
\cite{rikken93}; the dataset is found in the R package \rpkg{sp}.
Figure~\ref{fig:maas} plots the interval spacing of the density of
organic matter.  Dots at the individual spacings have discrete levels
because the measurements were taken to one decimal point.  There is a
strong peak at index 144, corresponding to 13.0\%~soil/kg.  Its
probability is 0.049 in the longest runs test, 0.065 in the
permutation height test, and 0.002 from the bootstrap.  It represents
measurements taken along the river, within 50--80~m.  A second peak
at index 42 (5.0\%~soil/kg) passes the run height test with a
probability of 0.004, but has no clear interpretation in terms of the
distance from the  river; the run length and bootstrap tests are not
significant, with probabilities of 0.322 and 0.969.  The interval
spacing contains two flats, one between indices 47 and 95
(5.3--7.2\%~soil/kg), the other between 97 and 125 (7.3--9.2\%~soil/kg).
Both have bootstrap probabilities of 0.000.  The spacing within the
flats is different, which can indicate a change in the scale parameter
of the underlying distribution.  A Mood test confirms a change in the
variance at 300~m from the river.

\bibliographystyle{IEEEtran}
\bibliography{IEEEabrv,dmodal}

@preamble{ " \newcommand{\noop}[1]{} " }

@UNPUBLISHED{kreider25c,
  author = "Greg Kreider",
	title = "Modality Analysis via Spacing with the {D}imodal Software Libraries",
	year = "2025",
  note = "https://arxiv.org/abs/2607.06722"
}

@TECHREPORT{rikken93,
  author = "M. G. J. Rikken and R. P. G. van Rijn",
  title = "Soil pollution with heavy metals --- an inquiry into spatial
    variation, cost of mapping and the risk evaluation of copper, cadmium, 
    lead and zinc in the floodplains of the {M}euse west of {S}tein, the
    {N}etherlands",
  year = 1993,
  institution = "Utrecht University",
  address = "Department of Physical Geography",
  note = "Doctoraalveldwerkverslag"
}

@ARTICLE{anderson57,
  author = "T. W. Anderson and Leo A. Goodman",
  title = "Statistical Inference About {M}arkov Chains",
  journal = "The Annals of Mathematical Statistics",
  volume = 28,
  number = 1,
  year = 1957,
  month = mar,
  pages = "89--110"
}

@ARTICLE{chung76,
  author="Kai Lai Chung",
  title = "Excursions in {B}rownian Motion",
  journal = "Arkiv f{\"o}r Matematik",
  volume = 14,
  number = "1--2",
  year = 1976,
  month = dec,
  pages = "155--177"
}

@INPROCEEDINGS{dorea14,
  author = "Chang C. Y. Dorea and Catia R. Goncalves and Paulo A. A. Resende",
  title = "Simulation Results for {M}arkov Model Selection: {AIC}, {BIC} and {EDC}",
  booktitle = "Proceedings of the World Congress on Engineering and Computer
               Science",
  volume = "II",
  organization = "WCECS",
  address = "San Francisco",
  year = 2014,
  month = oct
}

@ARTICLE{fu94,
  author = "J. C. Fu and M. V. Koutras",
  title = "Distribution Theory of Runs: A {M}arkov Chain Approach",
  journal = "Journal of the American Statistical Association",
  volume = 89,
  number = 427,
  year = 1994,
  month = sep,
  pages = "1050--1058"
}

@ARTICLE{hall97,
  author = "W. J. Hall",
  title = "The Distribution of {B}rownian Motion on Linear Stopping Boundaries",
  journal = "Sequential Analysis",
  volume = 16,
  number = 4,
  year = 1997,
  pages = "345--352"
}

@BOOK{helstrom95,
  author = "Carl W. Helstrom",
  title = "Elements of Signal Detection and Estimation",
  publisher = "PTR Prentice Hall",
  year = 1995,
  address = "Englewood Cliffs, New Jersey"
}

@ARTICLE{iafrate19,
  author = "F. Iafrate and E. Orsingher",
  title = "Some Results on The {B}rownian Meander With Drift",
  journal = "Journal of Theoretical Probability",
  volume = 33,
  year = 2019,
  pages = "1034--1060"
}

@ARTICLE{kaplansky45,
  author = "Irving Kaplansky and John Riordan",
  title = "Multiple Matching and Runs By The Symbolic Method",
  journal = "The Annals of Mathematical Statistics",
  volume = 16,
  year = 1945,
  pages = "272--277"
}

@ARTICLE{ma11,
  author = "Yanyuan Ma and Marc G. Genton and Emanuel Parzen",
  title = "Asymptotic properties of sample quantiles of discrete distributions",
  journal = "Annals of the Institute of Statistical Mathematics",
  volume = 63,
  year = 2011,
  pages = "227--243"
}

@ARTICLE{mood40,
  author = "A. M. Mood",
  title = "The Distribution Theory of Runs",
  journal = "The Annals of Mathematical Statistics",
  volume = 11,
  year = 1940,
  pages = "367--392"
}

@ARTICLE{philippou86,
  author = "Andreas N. Philippou and Frosso S. Makri",
  title = "Successes, Runs, and Longest Runs",
  journal = "Statistics and Probability Letters",
  volume = 4,
  year = 1986,
  month = jun,
  pages = "211--215"
}

@ARTICLE{shaughnessy81,
  author = "Peter W. Shaughnessy",
  title = "Multiple Runs Distributions: Recurrences and Critical Values",
  journal = "Journal of the American Statistical Association",
  volume = 76,
  number = 375,
  year = 1981,
  month = sep,
  pages = "732--736"
}

@ARTICLE{wald40,
  author = "A. Wald and J. Wolfowitz",
  title = "On A Test Whether Two Samples Are From The Same Population",
  journal = "The Annals of Mathematical Statistics",
  volume = 11,
  year = 1940,
  pages = "147--162"
}

@TECHREPORT{wallis41tr,
  author = "W. Allen Wallis and Geoffrey H. Moore",
  title = "A Significance Test for Time Series and Other Ordered Observations",
  institution = "National Bureau of Economic Research",
  address = "New York",
  month = sep,
  year = 1941,
  pages = "1--59"
}

@ARTICLE{wolfowitz43,
  author = "J. Wolfowitz",
  title = "On The Theory of Runs With Some Applications To Quality Control",
  journal = "The Annals of Mathematical Statistics",
  volume = 14,
  number = 3,
  year = 1943,
  month = sep,
  pages = "280--288"
}

\end{document}